\documentclass[a4paper]{article}
\usepackage{amsmath,amssymb,amsthm,amssymb,graphicx}
\usepackage[T1]{fontenc}
\usepackage[latin1]{inputenc}
\title{The role of von Neumann and L\"uders postulates in the
EPR-Bohm-Bell considerations: \\ Did EPR make a mistake?}
\author{Andrei Khrennikov\\
School of Mathematics and Systems Engineering\\
University of V\"axj\"o, S-35195, Sweden}

\begin{document}

\maketitle

\abstract{We show that the projection postulate plays a crucial
role in the discussion on the so called "quantum nonlocality", in
particular in the EPR-argument. We stress that the original von
Neumann projection postulate was crucially modified by extending
it to observables with degenerate spectra (the L\"uders postulate)
and we show that this modification is highly questionable from a
physical point of view, and is the real source of "quantum
nonlocality". The use of the original von Neumann postulate
eliminates this problem: instead of an [action at a
distance]-nonlocality we obtain a classical measurement
nonlocality, which is related to the synchronization of two
measurements (on the two parts of a composite system). It seems
that Einstein-Podolsky-Rosen did mistake in their 1935-paper: if
one uses correctly von Neumann projection postulate,  no
``elements of reality'' can be assigned to entangled systems. Our
analysis of the EPR and projection postulate makes clearer Bohr's
considerations in his reply to Einstein.}

\section{Introduction}

We shall show that the main source of debate in the EPR argument
\cite{EPR}, as well as in the discussion on the EPR-Bohm and
Bell's inequality \cite{B0,B}, is the misuse of von Neumann's
projection postulate \cite{VN}.

The projection postulate (PP) plays indeed a crucial role in the
EPR argument \cite{EPR}. We consider a
 composite system $s=(s_1,
s_2)$. Assigning an element of reality to $s_2$ on the basis of a
measurement performed on $s_1$ is fundamentally based on the PP.
Although von Neumann had presented strong physical arguments
stressing that the PP should be applied only to observables with
nondegenerate spectra \cite{VN}, it was nevertheless applied by
Einstein, Podolsky and Rosen \cite{EPR} in the case of observables
with degenerate spectra. This misapplication of von Neumann's
projection postulate \cite{VN} was later on formalized as a custom
by L\"uders \cite{LUDER}.

Hence, we show that the EPR-argument is in fact based on an
improper extension of von Neumann's postulate, and that as such it
should not be considered as a valid attack against the Copenhagen
interpretation (which is itself based solely on von Neumann's
axiomatic \cite{VN}). It means that Quantum Mechanics can (but, in
principle, need not) be interpreted as a complete and local
theory.

Our analysis shows that measurements on entangled systems, as
predicted by QM formalism, are nonlocal. However, such a
measurement nonlocality is essentially a classical synchronization
nonlocality and has nothing to do with action-at-a-distance. Thus,
instead of the ``state-nonlocality,'' everything is reduced to
nonlocality of the (classical) design of the EPR-experiment.

The presence of a measurement nonlocality in the EPR-Bohm
experiment implies the violation of the conditions of Bell's
theorem and, hence, provides a possibility to violate Bell's
inequality,  cf. works of Hess and Philipp on the role of the time
parameter in the EPR-Bell framework, \cite{HP}--\cite{HPX}.

We remind that the projection postulate is nowadays formulated in
the following form:

\medskip

{\bf PP}:  {\it Let $\; a\; $ be a physical observable
described by a self-adjoint operator $\widehat{a}$ having purely
discrete spectrum. Any measurement of the observable $\; a\; $ on
the pure quantum state $\psi$ induces a transition from the state
$\psi$ into one of the eigenvectors $e_a^k$ of the operator
$\widehat{a}.$}

\medskip

It is in this form that the projection postulate
was used by EPR in \cite{EPR} as well as in numerous discussions
on "quantum nonlocality."

Recently I submitted as arXiv-preprint another paper on the role
of the projection postulate in the EPR-argument, see \cite{UUU}.
This paper contains a detailed (step by step) analysis of the
EPR-considerations. I should do this, because I received a lot of
critical comments on the present paper in which it was claimed
that the projection postulate did not play an important role in
the EPR-paper. Another recent preprint \cite{UUU1} is devoted to
the role of von Neumann's projection postulate in quantum
teleportation and main quantum algorithms. The main conclusion of
the latter paper is that quantum teleportation would not provide
the expected result (if one keeps to the orthodox Copenhagen).
Roughtly speaking quantum teleportation is an artifact of the
misuse of von Neumann's postulate. On the other hand, quantum
information schemes incorporated in known quantum algorithms are
consistent with von Neumann's postulate.

\section{Quantum description of measurements on composite systems}

Let $H_1$ and $H_2$ be two complex finite dimensional Hilbert
spaces, $\dim H_i \geq 2.$ Let $\widehat a_1: H_1 \to H_1$ and
$\widehat a_2: H_2 \to H_2$ be two self-adjoint operators. The
Hilbert space $H_i$ represents (quantum) states of the system
$s_i, i= 1, 2.$ The operator $\widehat a_i$ represents an
observable $a_i$ corresponding to measurements on $s_i, i = 1, 2.$
The composite system $s=(s_1, s_2)$ is described by the tensor
product space $H=H_1 \otimes H_2.$

The operators $\widehat A_1= \widehat a_1 \otimes I$ and $\widehat
A_2= I \otimes \widehat a_2$ represent partial measurements $A_1$
and $A_2$ on $s: a_1$ on $s_1$ and $a_2$ on $s_2,$ respectively.

To simplify considerations, we assume that both operators
$\widehat a_1$ and $\widehat a_2$ have purely discrete
nondegenerate\footnote{The latter condition is redundant. We would
like just to emphasize that degeneration of spectra in $H_i$ does
not play any role. Even operators with nondegenerate spectra in
$H_i$ induce operators with degenerate spectra in $H.$} spectra.
We consider eigenvectors $e_1^\alpha$ and $e_2^\beta$ of these
operators: $\widehat a_1 e_1^\alpha= \lambda_1^\alpha e_1^\alpha,
\alpha=1, \ldots, N_1=\dim H_1$ and $\widehat a_2 e_2^\beta=
\lambda_2^\beta e_2^\beta, \beta=1, \ldots, N_2=\dim H_2.$

By taking two arbitrary (independent) eigenvectors for each
operator we construct an "entangled state":
\begin{equation}
\label{ETS} \psi= c_1 e_1^i \otimes e_2^j + c_2 e_1^j \otimes
e_2^i, |c_1|^2 + |c_2|^2=1.
\end{equation}
Suppose a measurement of $A_1$ was performed  on the composite
system $s=(s_1, s_2),$ i.e., $a_1$ was performed on $s_1.$ Suppose
that the result $$A_1=\lambda_1^i$$  was obtained. This measurement
is represented by the operator $\widehat A_1= \widehat a_1 \otimes
I.$

By the projection postulate PP the state $\psi$ is projected onto
$$\psi_1^{ij}= e_1^i \otimes e_2^j.$$
Thus instantaneously (and in principle without any interaction of
the $A_1$-measurement device with the system $s_2$) the state of
$s_2$ is changed. It became sharply determined: $\widehat A_2
\psi_1^{ij}= \lambda_2^j \psi_1^{ij},$ and hence $A_2=
\lambda_2^j.$

This is nothing else than the so called "quantum nonlocality". To
get it one need not appeal to Bell's inequality \cite{B0},
\cite{B} (and hidden variables at all).\footnote{I remark that
precisely in this way the EPR-Bohm experiment was presented by
Alain Aspect in his talk at the V\"axj\"o conference \cite{KHR}:
by measuring polarization of $s_1,$ one projects the state of the
composite system and makes the state of $s_2$ determined, see also
\cite{AA}. } If one takes this into account Bell's considerations
would play a subsidiary  role. The main problem is to explain
"quantum nonlocality" as it follows from the quantum formalism.

\section{Abuse of von Neumann's projection postulate}
By reading von Neumann's book \cite{VN} I found that the modern
formulation PP of the projection postulate, see introduction, is
not the original von Neumann's formulation at all. One extremely
important condition is omited. It is the condition of {\it non
degeneration of spectrum} of the quantum observable. The
formulation PP is, in fact, L\"uders' postulate \cite{LUDER} and
not at all von Neumann's one \cite{VN}.

Opposite to L\"uders, von Neumann sharply separated the cases of
nondegenerate and degenerate spectra. The PP could be applied only
in the first case. In the second case the result $a=\lambda$ does
not determine any definite state. To obtain a definite state, one
should perform a refinement $d$ of the $a$-measurement, such that
$a=f(d)$ and $d$ is represented by an operator $\widehat d$ having
nondegenerate spectrum.

Let us go back to the situation described in section 2. Here the
operators $\widehat A_1$ and $\widehat A_2$ (corresponding to
measurements on $s_1$ and $s_2)$ always have degenerate spectra:

If e.g. $\widehat a_1 e=\lambda e, e \in H_1,$ then $\widehat A_1
\psi=\lambda \psi$ for any $\psi=e \otimes \phi \in H, \phi \in
H_2.$ (We just remind that $\dim H_2 \geq 2$).

By von Neumann the result $A_1= \lambda_1^j$ does not induce
projection onto a definite pure state. The state of the system
$s=(s_1, s_2)$ is not determined after such a {\it partial}
measurement!

Thus if one follows really  the  Copenhagen
interpretation,   no trace of  "quantum nonlocality" would be found.

\section{Von Neumann's postulate and "classical nonlocality"}

What would von Neumann recommend to do to get the definite post
measurement state? He would recommend to perform a refinement $A$
of the $A_1$-measurement which would be represented by an
operator, say $\widehat A,$ having nondegenerate spectrum.

The crucial point is that one could not construct such a
refinement $\widehat A$ by operating only in $H_1,$ i.e., by using
operators of the form
$$
\widehat A= \widehat C \otimes I, \widehat C: H_1 \to H_1.
$$
One should consider "nonlocal measurements" which are represented
by operators acting in the complete tensor product $H= H_1 \otimes
H_2.$

In particular, one can not create a nondegenerate refinement of
the $s_1$-spin measurement via modification of the spin operator
in $H_1={\bf C}^2.$ The corresponding nongenerate operator is
nontrivial in the whole $H={\bf C}^4.$

The simplest refinement $\widehat A$ can be constructed as
$\widehat A e_1^\alpha \otimes e_2^\beta= \gamma_{\alpha\beta}
e_1^\alpha \otimes e_2^\beta, $ where $\gamma_{\alpha\beta} \ne
\gamma_{\alpha^\prime \beta^\prime}, $ if $\alpha \ne
\alpha^\prime$ or $\beta \ne \beta^\prime.$

\medskip

{\bf Example.} (Spin refinement) Let $\widehat a_1= \sum^3_{i=1}
x_i \sigma_i$ and $\widehat a_2= \sum^3_{i=1} y_i \sigma_i,$ where
$\sigma_i$ are Pauli matrices. Here $H_1=H_2= {\bf C}^2,$ hence,
$H={\bf C}^4.$ Consider eigenvectors $\widehat a_i e_i^{\pm} = \pm
e_i^{\pm}, i = 1,2.$ We consider the following encoding: $-- = 0 0
= 0, +- = 10 = 1, -+ = 01 =2, ++ = 11=3.$ We now set $\widehat A
e_1^\alpha \otimes e_2^\beta = \alpha \beta e_1^\alpha \otimes
e_2^\beta.$ This operator has nondegenerate spectrum $\lambda=
0,1,2,3.$ We have $\widehat A_1= f_1 (\widehat A),$ where $f_1
(\alpha \beta)=\alpha.$ In the same way $\widehat A_2=f_2
(\widehat A),$ where $f_2(\alpha \beta)= \beta.$ By von Neumann
only measurement of the observable $A$ represented by $\widehat A$
induces projection of the entangled state $\psi$ onto the pure
state $e_1^\alpha \otimes e_2^\beta$ (in the  case of the result
$\lambda= \alpha \beta).$ Thus, instead of mysterious ``quantum
nonlocality'' (or state nonlocality), we have measurement
nonlocality which is purely classical nonlocality.

\medskip

Measurement which is performed on a composite system $s=(s_1,
s_2)$ consisting of two spatially separated parts is, of course,
nonlocal. It is not surprising that it is represented by a
"nonlocal operator" $\widehat A,$ cf. \cite{KZ}. Nothing nonlocal
happens with the state of $s.$ Only (classical) design of the
experiment is nonlocal.

\section{Simultaneous measurement of two compatible observables}

I expect that my previous arguments on classical measurement
nonlocality in experiments with entangled systems would be
criticized in the following way: "There is nothing about
measurement nonlocality, since one can perform, instead of
measurement of a nonlocal observable $A,$ simultaneous
measurements of two compatible observables $A_1$ and $A_2.$ The
crucial point is that the operators $\widehat A_1$ and $\widehat
A_2$ commute."

Such an argument would be based on another von Neumann postulate,
namely about measurement of compatible observables \cite{VN}, p.
200-201:

\medskip
(PC) {\it The probability that in the state $\psi$ the quantities
with (commuting)\footnote{"Commuting" was absent in the original
postulate. Von Neumann formulated (PC) first for arbitrary
self-adjoint operators, but then after analyzing it he pointed out
that they should commute.}  operators $\widehat R_1, \ldots,
\widehat R_n$ take on values from respective intervals $\Delta_1,
\ldots, \Delta_n$ is
$$||E_1 (\Delta_1) \ldots E_n (\Delta_n) \psi ||^2,$$
where $E_1 (\lambda), \ldots, E_n (\lambda)$ are resolutions of
the identity belonging to $\widehat R_1, \ldots, \widehat R_n,$
respectively.}

\medskip

One may say that in measurements for entangled states one can
consider simultaneous measurement of two compatible observables
represented by the operators $\widehat A_1= \widehat a_1 \otimes
I$ and $A_2= I \otimes \widehat a_2, [\widehat A_1, \widehat
A_2]=0.$ Hence the story is about simultaneous measurement of
compatible observables and not about measurement of ``nonlocal''
observable  $A$ represented by $\widehat A.$

Again by reading von Neumann \cite{VN}, p. 201-206, we understand
that the above argument does not take into account the crucial
fact that {\it simultaneous measurement of two compatible
observables is not reduced to separate measurement of each of
them.} By the conventional quantum formalism simultaneous
measurement of $a$ and $b$ given by $\widehat a$ and $\widehat b,
[\widehat a, \widehat b]=0,$ is performed in the following way.

One should construct an observable $d$ (having nondegenerate
spectrum)  represented by $\widehat d$ such that $\widehat a= f_1
(\widehat d)$ and $\widehat b= f_2 (\widehat d).$ Then
simultaneous measurement of $a$ and $b$ is performed in two steps:

\medskip

a) measurement of $d$;

b) $a$ and $b$ are obtained as $a=f_1 (d)$ and $b=f_2 (d).$

\medskip

Therefore in the case of measurements on composite systems,
$s=(s_1, s_2),$ one could not (!) proceed without (classically)
nonlocal  measurement of a refinement $A$ given by $\widehat A.$

\medskip

What does it mean physically?

\medskip

It means that one should make synchronized measurement on both
systems. Then the result $\lambda=\alpha \beta$ should be decoded
into $A_1=\alpha, A_2=\beta.$ After this one can calculate e.g.
the correlation $<A_1 A_2>$ between $A_1$ and $A_2.$ Nonlocality
appears here via synchronization. An observer to whom such a
synchronization is not available would not be able to find the
right matching between the results of $A_1$ and
$A_2$-measurements.

Thus the EPR-Bohm experiment is really nonlocal, but it is
measurement nonlocally which is completely classical time
synchronization nonlocality.

\section{The original EPR experiment}
It is evident that Alain Aspect simply borrowed  "quantum
nonlocality" argument \cite{AA} from  EPR. Thus the root of
misunderstanding was in the original paper \cite{EPR}.

Let now $H1=H_2= L_2 (R^3, dx).$ Let $a_1$ and $a_2$ be
observables represented by operators $\widehat a_1$ and $\widehat
a_2$ with purely discrete nongenerate spectra:
$$\widehat a_i e_i^\alpha = \lambda_i^\alpha e_i^\alpha, i=1,2.$$
Any state $\psi \in H= H_1\otimes H_2$ can be represented as
$$\psi=\sum_{\alpha, \beta} c_{\alpha \beta} e_1^\alpha \otimes e_2^\beta,$$
where $\sum_{\alpha, \beta}|c_{\alpha \beta}|^2=1.$ Einstein,
Podolsky and Rosen claimed that measurement of $A_1$ given by
$\widehat A_1=\widehat a_1 \otimes I$ induces projection of $\psi$
onto one of states $e_1^\alpha \otimes u, u \in H_2.$ In
particular, for a state of the form $\psi=\sum_\gamma c_\gamma
e_1^\gamma \otimes e_2^\gamma$ one of states $e_1^\gamma \otimes
e_2^\gamma$ would be realized. Thus by performing measurement on
the $s_1$ with the result $\lambda_1^\gamma$ the "element of
reality $a_2=\lambda_2^\gamma$" is assigned to $s_2$.

However, the EPR considerations did not match von Neumann's
projection postulate, because the spectrum of $\widehat A_1$ is
degenerate. Finally, (after consideration of operators with
discrete spectra), Einstein, Podolsky and Rosen considered
operators of position and momentum having continuous spectra.
According to von Neumann \cite{VN} one should proceed by
approximating operators with continuous spectra by operators with
discrete spectra.

Thus by von Neumann to get "an element of reality" one should
perform measurement of ``nonlocal observable'' $A$ given by a
nonlocal refinement of e.g. $\widehat A_1= \widehat q_1 \otimes I$
and $\widehat A_2= I \otimes \widehat p_2.$

We point out that von Neumann's viewpoint coincides with Bohr's
viewpoint in his reply to Einstein \cite{BR}. Unfortunately, Bohr
did not use mathematical arguments of von Neumann \cite{VN} to
explain to Einstein why to produce "an element of reality"  for
the subsystem $s_2$ one should measure $A$ on $s=(s_1, s_2)$ and
not just e.g. $A_1$ on $s_1.$

\section{Bell's inequality}

What are consequences of our analysis of the EPR-Bohm and the EPR
experiments for Bell's considerations?

\subsection{Measurement nonlocality as opposed to action at the
distance}

 The main consequence is that classical random variables
(if they exist) for $A_1$ and $A_2$ measurements should have the
form $A_1 (\lambda)= f_1 (d(\lambda)), A_2(\lambda)=f_2
(d(\lambda)),$ where $d(\lambda)$ is the variable corresponding to
synchronized measurement on both systems. Since $d \equiv
d_{A_1,A_2},$ Bell's condition of locality is violated. Therefore
it is nothing surprising that Bell's inequality can be violated.

The crucial point is that such {\it measurement nonlocality} has
nothing to do with action at the distance. It is classical time
synchronization nonlocality, cf. works of Hess and Philipp on the
role of the time parameter in the EPR-Bell framework,
\cite{HP}--\cite{HPX}. The authors of these papers rightly
stressed the role of analysis of time synchronization in
measurements on pairs of entangled photons.

\subsection{Time synchronization and time window}
The time synchronization viewpoint in the EPR-Bohm experimentmeans
that experimenters should couple in time two clicks of detectors
corresponding to measurements on the first and second subsystems,
respectively. Of course, such a time coupling is nonlocal. Roughly
speaking if Alice and Bob count the first and the second types of
clicks, respectively, they should call each other to couple their
clicks in time. This procedure has nothing to do with QM. It is a
classical nonlocal design of the experiment.

\medskip

Which consequences for studies on Bell's inequality has this
viewpoint to the EPR-Bohm experiment?

\medskip

First of all, the time synchronization viewpoint to the problem of
locality (based on the correct application of von Neumann's
projection postulate) stressed the role of time coupling in the
EPR-Bohm experiment. Consider following settings of polarization
beam splitters in the EPR-Bohm experiment: 1) $a$  for
measurements on the first subsystem; 2). $b$ for measurements on
the second subsystem. Suppose that one measures the pair of
observables which are represented by operators $\widehat
A=\widehat a \otimes I$ and $\widehat B= I\otimes \widehat b.$ In
the real experiment one could not expect the detectors would click
simultaneously. The first and the second photons can have
different delays induced by passing through polarization beam
splitters and  electro-optic modulators, see \cite{WWW} for
details and references. Moreover, I would like to point to another
source of delays. In modern experiments a pair of entangled
photons is produced via interaction of a laser pulse with the
crystal. It is important for us that photons are emitted not
simultaneously. Thus delays can appear from the very beginning.

Thus in the real EPR-Bohm experiment (in opposite the ideal one)
time synchronization is more complicated. It is not the
coincidence time synchronization, but synchronization by using the
time window, see \cite{WWW} for details.

Let $A$-measurements and $B$-measurement produce clicks at the
moments:
$$
t_1^a, t_2^a,..., t_N^a,
$$
$$
t_1^b, t_2^b,..., t_M^b
$$
(in general $N\not=M).$

Suppose that a time window $\Delta$ is fixed. Then two clicks are
coupled to the same measurement iff
\begin{equation}
\label{CL} \vert t_j^a - t_j^b\vert \leq \Delta.
\end{equation}
This condition is definitely nonlocal! Hence, in  the real
EPR-Bohm experiment measurement nonlocality via time
synchronization is even more evident than in the ideal experiment.

Can one simulate numerically  the  EPR-Bohm correlations by using
time synchronization argument?

The answer is to be positive, see  \cite{Raedt}-- \cite{Raedt2}.

In principle, one may expect that it would be done, since the time
window condition (\ref{CL}) induces unfair sampling, see \cite{}.
Since some pairs of clicks violate (\ref{CL}),  they are not taken
into account. If one were to choose another setting for the second
polarization beam splitter, say $b^\prime,$  then another
collections of pairs of clicks and hence pairs of photons would be
selected via the time window condition:
\begin{equation}
\label{CL} \vert t_j^a - t_j^{b^\prime} \vert \leq \Delta.
\end{equation}
Of course, design of a natural algorithm essentially supported
this point,  \cite{Raedt}-- \cite{Raedt2}.

\subsection{Frequency approach to the EPR-Bohm experiment}

The time synchronization consequence of our analysis of the role
of von Neumann's postulate for the EPR-Bohm experiment also
supports the frequency (von Mises)  approach to Bell's inequality,
see \cite{KHI}-\cite{KHFR}. It is very natural to consider
collectives (random sequences) given by sequences of pairs of
clicks which are selected via the time window  condition. By
operating with frequency probability, instead of measure-theoretic
probability (which was used by Bell \cite{B}: it was denoted by $d
\rho(\lambda)),$ I obtain the EPR-Bohm correlations in the local
realistic (in  the sense of absence of action at the distance)
framework, see  \cite{KHI}-\cite{KHFR}.

\subsection{Probabilistic (in)compatibility}

On the other hand, in the presence of time synchronization it is
extremely unnatural to assume as Bell did that all observables
(measured in a few incompatible experiments) can be represented by
random variables on the same probability space: $a(\lambda),
b(\lambda), b^\prime(\lambda),...$ (so one can use the unique
probability measure $d \rho(\lambda))$ which does not depend on
experimental settings). The latter assumption was called in
\cite{KHPC} probabilistic compatibility assumption - PC. Its role
in Bell's argument was studied in detail \cite{KHPC}. In
particular, it was pointed out that violation of PC (and hence
Bell's inequality) has been studied in probability theory and
statistics during the last hundred years.

\subsection{Contextuality}

Finally, we remark about contextuality. As was pointed by Bell
\cite{B}, contextuality -- in the sense of taking into account
simultaneous measurements of compatible observables -- blocks
derivations of Bell-type inequalities, see S. Gudder \cite{GU} for
deep analysis of relation between contextuality and Bell's
theorem. The crucial point is that Bell personally considered
nonlocality as the state nonlocalty and not the measurement one.
Therefore contextuality was explained in such a way: it can be
generated by nonlocal state reduction.  In our approach
contextuality can be naturally explained by classical (time
synchronization) measurement nonlocality.

\section{Lessons}

The main lesson of our considerations (which is especially
important for students) is that one should start with reading of
original sources (such as \cite{EPR}, \cite{BR}, \cite{VN},
\cite{LUDER}), even if such sources are considered as difficult
for reading. Unfortunately, deepness of investigations is often
interpreted as unclearness of presentation. The book of von
Neumann \cite{VN} still provides the deepest analysis of quantum
foundations (even comparing with the most advanced modern books).

\section{Did EPR make a mistake?}

Recently Karl Hess wrote in his Email-comment on this preprint:
"Did EPR make a mistake? I was sure that they proved
incompleteness of quantum mechanics."

The answer is not so simple, since the whole EPR-story is very
complicated. In fact, this story is not about QM by itself, but
about its interpretations. EPR wanted to show that QM endowed with
the Copenhagen interpretation is not complete. It is crucial that
they did not claim that QM endowed with any interpretation is
incomplete. The main problem is that EPR did not formulate
precisely the interpretation under attack! It was more or less
clear that it was the interpretation of Bohr and Heisenberg.

As we have already pointed out, EPR's argument was heavily based
on the projection postulate. They definitely assigned it to the
criticized interpretation. The possibility to apply the projection
postulate for operators having degenerate spectra played the
fundamental role in the EPR-considerations.

However, one might be curious: "Would fathers of the Copenhagen
interpretation accept such a (mis?) use of the projection
postulate?" Shortly the question is "Was EPR's argument against
the real Bohr-Heisenberg interpretation? May be it was simply
based on EPR's misinterpretation of views of Bohr and Heisenberg?"

I remark that in this framework it would be better to speak about
views of concrete persons, since the "Copenhagen interpretation"
is an extremely diffuse collection of views, see Arcady Plotnitsky
\cite{PL} for details. I have not yet studied well views of
Heisenberg.  But I read  Bohr's reply \cite{BR}  to Einstein. I
have the impression that Bohr wrote to Einstein (unfortunately, in
nonmathematical terms) that no "element of reality" can be
"created" by measurement on a single sub-system. To create an
"element of reality" one should perform measurement on the second
sub-system. This is nothing else than von Neumann's refinement
measurement!

I would say that in 1935 EPR misinterpreted the Copenhagen
interpretation (at least the Bohr-von Neumann one). They attacked
a sort of "perverse Copenhagen interpretation" based on abuse of
the von Neumann projection postulate.\footnote{Von Neumann's book
was being on the book-shell in Einstein's office. Did Einstein
read it?} They demonstrated that QM endowed with this EPR-version
of the Copenhagen interpretation was incomplete.\footnote{So
called quantum nonlocality was considered as totally absurd at
that time.} However, the EPR-argument did not imply that QM with
the real (Bohr-von Neumann) Copenhagen interpretation was
incomplete.

Unfortunately, the situation was not clarified at the very
beginning. Bohr's reply was not sufficiently clear and it was not
coupled to von Neumann's book.  And later the EPR-story developed
in really unexpected way. The pseudo-Copenhagen interpretation of
EPR became  the orthodox Copenhagen interpretation. In particular,
Bohr's reply and hence his views were ignored. Nowadays  L\"uders'
postulate is widely used,  instead of von Neumann's projection
postulate. Thus the EPR arguments became fundamentally important
and their stimulated Bell toward his inequality argument. He
definitely believed that QM is not complete (as it was "proved" by
EPR). And incompleteness was understood in the EPR-fashion.

One may ask: "Is it the end of the hidden variable story?" Not at
all. It is just a new beginning. Instead of the EPR-incompleteness
and assigning to a hidden variable $\lambda$ values of quantum
observables, $\lambda \to a(\lambda),$ we can analyze
possibilities of more complicated couplings between prequantum and
quantum worlds, see e.g. \cite{KJ}.

{\bf Conclusion.} {\it The EPR paradox is a consequence of
misinterpretation of the Copenhagen interpretation based on the
vague application of von Neumann's projection postulate.}

\section{Appendix: Projection postulate}

After discussions with a few hundred scientists working in quantum
foundations and quantum information (who visited my university
during a series of V\"axj\"o conferences on quantum foundations)
it became completely clear that the majority has even no idea
about the difference between von Neumann's and L\"uders'
projection postulates (PPs). Those who knew the story about  PP
expressed their strong opinion that von Neumann's book is old
fashioned and  at the very beginning many things were not
sufficiently clear. Later a lot in quantum foundations was
clarified. In particular, L\"uders clarified PP which was
formulated by von Neumann in the very complicated form. I strongly
disagree with such a common opinion. The story is not about
simplification of von Neumann's arguments. It is about
misunderstanding of the basis postulate of QM. As was already
pointed out, the root of L\"uders' misunderstanding was in similar
misunderstanding of    PP by Einstein, Podolsky and Rosen in their
fundamental article \cite{EPR}. Therefore I find important to
present this appendix containing practically forgotten comparing
of von Neumann's PP and L\"uders' PP. Finally, we recall that this
comparison is extremely important not only for quantum
foundations, EPR-paradox, Bell's inequality. The difference
between two forms of PP should be pointed out even without any
relation to mentioned questions. We recall that the basic
definition of quantum conditional probability which is widely used
in quantum information theory is based on L\"uders PP and not on
von Neumann's one!

It should be emphasized that comparasion of two PPs is not of only theoretical value. In principle, it can be
tested experimentally: either the post-measurement state is pure (as L\"uders and quatum majority
claimed) or it is really not well defined. Of course, one should understand better the meaning of
``non well defined state.'' However, we prefer to come back to this problem in a coming article.

\subsection{L\"uder's Projection Postulate}

It is very important for our further considerations to remark that
if spectrum of an operator $\widehat{a}$ is {\it degenerate} then
(according to von Neumann \cite{VN}) the observation of the result
$a=\alpha_k$ {\it does not induce the transition of the initial
pure state $\psi$ into a new pure state.} In such a case the
resulting state is not determined. It can be determined only
through a subsequent measurement of an observable $d$ refining the
original observable $a.$ After such a refining measurement we
obtain not a pure state, but a statistical mixture.

However, in the contemporary formulations of ``the von Neumann
projection postulate'' the cases of nondegenerate and degenerate
spectra are not distinguished! In fact, this was not the original
von Neumann invention, but it was a new postulate proposed by G.
L\"uders in 1951,  \cite{LUDER} :

\medskip

{\bf L\"uders Projection Postulate.} {\it For  any operator
$\widehat{a}$ with purely discrete spectrum, a measurement  of the
corresponding observable $a$ giving the result $a= \alpha,$ where
$\alpha \in \rm{Spec}(\widehat{a}),$ always
 produces the projection
of the initial state $\psi$ onto the state
\begin{equation}
\label{LUI}
 \psi_\alpha= \frac{P_{\alpha} \psi}{\Vert P_{\alpha}
\psi \Vert},
\end{equation}
where as always $P_{\alpha}$ is the orthogonal projector onto the
subspace corresponding to the eigenvalue $\alpha.$}

\medskip
The crucial point is that by the L\"uders' PP the
post-measurement state is also a pure state independently of
degeneration of spectrum.

However, von Neumann emphasized a few times in \cite{VN} that
{\small `` if the eigenvalue $\alpha$ is multiple, then the state
$\phi$ after the measurement is not uniquely determined by the
knowledge of the result of the measurement,''} p. 218.

 The post-measurement state  $\phi$ is not determined. What does it mean?
J. von Neumann pointed out that to determine $\phi$ one should
determine a subsequent measurement procedure which corresponds to
the choice of a concrete orthonormal basis in the subspace
$$
{\cal H}_\alpha = P_{\alpha}^a {\cal H}.
$$
This ambiguity in the determination of the post-measurement state
is an important difficulty in foundations of quantum
mechanics.\footnote{Let us take this problem seriously. It might
be interpreted as a sign of  {\it incompleteness of quantum
mechanics:} the quantum formalism  does not determine the
post-measurement state in the case of observables represented by
operators with degenerate spectra. However, von Neumann did not
interpret this problem in such a way. It may be that success of
L\"uders' modification of the von Neumann projection postulate is
due to common wish to escape this problem.}

\subsection{Von Neumann's and L\"uders' postulates for mixed states}

Let $\psi$ be a pure state and let $P$ be a projector. By
L\"uders' postulate after measurement of the observable
represented by $P$ that gives the result $1$ the initial pure
state $\psi$ is transformed again into a pure state, namely,
$$
\psi^\prime= \frac{P\psi}{\Vert P\psi \Vert}.
$$
Thus for corresponding density operators we have:
$$
\rho_{\psi^\prime}= \psi^\prime \otimes \psi^\prime= \frac{P \psi
\otimes P\psi}{\Vert P\psi\Vert^2}.
$$
We remark that $(P \psi \otimes P\psi) \phi= \langle \phi,P\psi \rangle  P\psi$ and
that $(P \rho_{\psi} P) \phi= \langle P\phi,\psi \rangle  P\psi.$ As $P^\star=P,$
we obtain that
\begin{equation}
\label{LVN} \rho_{\psi^\prime}= \frac{P \rho_{\psi}P}{\Vert
P\psi\Vert^2}.
\end{equation}
Finally, we see that $\Vert P \psi \Vert^2= \langle  P \psi, P \psi \rangle = \langle  P^2
\psi, \psi \rangle = \langle  P \psi, \psi \rangle = \rm{Tr} \; \rho_\psi P.$ Thus:
\begin{equation}
\label{LVN1} \rho_{\psi^\prime}= \frac{P \rho_{\psi}P}{\rm{Tr}\;
\rho_\psi P}.
\end{equation}
In this way L\"uders' postulate is represented in the framework of
density operators (still in the case of pure states). Gerhart
L\"uders generalized this formula (without any doubt) to an
arbitrary state $\rho.$ If we measure the observable represented
by a projector $P$ for the ensemble of systems described by $\rho$
and then select a new ensemble corresponding to the result 1, we
get the state:
\begin{equation}
\label{LVN2} \rho^\prime= \frac{P \rho P}{\rm{Tr}\; \rho P}.
\end{equation}
Let now consider an arbitrary self-adjoint operator with purely
discrete spectrum:
\begin{equation}
\label{BBB} \widehat{a}=\sum_{m} \alpha_m P_m ,\; \alpha_m \in {\bf
R},\; P_m P_l = \delta_{ml} P_l.
\end{equation}
G. L\"uders pointed out \cite{LUDER} that in a measurement of
$\widehat{a}$ the initial state  $\rho$ is transformed into
\begin{equation}
\label{LVN3} \rho^\prime= \sum_m P_m \rho P_m .
\end{equation}
It is easy to see that such a $\rho^\prime$ is again positive and
self-adjoint and its trace equals to one.  This is  L\"uders' postulate for the transformation of a state
$\rho$ through a measurement of an observable represented by
$\widehat{a}.$ The crucial L\"uders' assumption is that, {\it for
a pure state after a measurement of $\widehat{a}$ and selection
with respect to the value $\alpha_m,$ we always obtain again a
pure state.}

\medskip

Von Neumann had the completely different viewpoint on such a
transformation \cite{VN}. As was already pointed out, even for a
pure state $\psi$ the result will not again be a pure state  (if
the operator has degenerated spectrum).

 Let $\widehat{a}$ have nondegenerate spectrum.  Thus all
$P_m$ are one dimensional projectors (onto eigenvectors $\{e_m\}$
of $\widehat{a}).$ Then by the von Neumann projection postulate a
measurement of $a$ giving the result $a= \alpha_m$ really induces
the projection of the original pure state $\psi$ onto $e_m^a.$ The
transformation of the density operator
 is given by:
\begin{equation}
\label{GH} \rho_{\psi^\prime}= \sum_m P_m \rho_{\psi}P_m
\end{equation}
(so in the nondegenerate case L\"uders' approach coincides with
von Neumann's one). Starting with an arbitrary initial state
$\rho$ we obtain the state:
\begin{equation}
\label{GH1} \rho^\prime= \sum_m P_m \rho P_m.
\end{equation}
We remark that, since all projectors are one dimensional, we have:
$(P_m \rho P_m) \phi= P_m( \langle \phi, e_m \rangle  \rho e_m) = \langle  \rho e_m,
e_m \rangle \langle \phi, e_m \rangle  e_m= \langle  \rho e_m, e_m \rangle  P_m \phi.$ Thus we can rewrite
(\ref{GH}) as von Neumann  wrote:
\begin{equation}
\label{GHZ} \rho^\prime= \sum_m \langle \rho e_m, e_m \rangle  P_m.
\end{equation}

Let us start with a pure state $\psi.$ If $\widehat{a}$ has
degenerate (discrete) spectrum, then according to von Neumann
\cite{VN} a measurement of $a$ giving the value $a=\alpha_m$  {\it
does not induce a projection of} $\psi.$  The result will not be a
pure state (in particular, not $\psi_m = P_m \psi).$ Moreover, the
resulting state is not determined. Only a subsequent measurement
of an observable $d$ such that $a=f(d)$ and $d$ is represented by
the operator $\widehat{d}$ with nondegenerate spectrum will
determine the final state.

Let $\widehat{a}= P$ be an orthogonal  projector onto a subspace
${\cal H}_0$ of the state space ${\cal H}.$ Let us choose in
${\cal H}_0$ an orthonormal basis $\{\phi_n\}.$ The basis
$\{\phi_n\}$ can be completed to an orthonormal basis in ${\cal
H}:\{\phi_n , \phi_l^\prime\}.$ Let us take two sequences of real
numbers $\{\gamma_n\},\{\gamma_n^\prime\}$ such that all numbers
are distinct. We define the corresponding self-adjoint operator
$\widehat{d}$ having eigenvectors $\{\phi_n , \phi_l^\prime\}$ and
eigenvalues $\{\gamma_n,\gamma_n^\prime\}:$
$$
\widehat{d}= \sum_n \gamma_n P_{\phi_n} +\sum_n \gamma_n^\prime
P_{\phi_n^\prime}.
$$
Its domain of definition is given by
$$
D(\widehat{d})= \{ \psi \in {\cal H}: \psi= \sum_n \gamma_n^2
\vert \langle \psi, \phi_n \rangle \vert^2 + \sum_n
(\gamma^\prime_n)^2 \vert \langle \psi, \phi^\prime_n \rangle
\vert^2 < \infty\}.
$$

 J. von Neumann postulated \cite{VN}  that   one can construct a physical
observable $d$ described by the operator $\widehat{d}.$
Measurement of $d$ can be considered as measurement of the
observable $a,$ because $a=f(d),$ where $f$ is some function such
that $f(\gamma_k)=1$ and $f(\gamma_k^\prime)=0.$ But the
$d$-measurement (without post-measurement selection with respect
to eigenvalues) produces the statistical mixture:
$$
\bar{\rho}= \sum_n \vert \langle \psi, \phi_n \rangle \vert^2 P_{\phi_n} + \sum_n
\vert \langle \psi, \phi_n^\prime \rangle \vert^2 P_{\phi_n^\prime}.
$$
Since we can choose  $\{\phi_n\}$ and $\{\phi_n^\prime\}$  in many
ways, by obtaining the result $a= \alpha_k$ we cannot determine
the post-measurement state.

If we start with an arbitrary state $\rho$ and an arbitrary
self-adjoint operator $\widehat{a}$ with purely discrete spectrum,
then we can determine the post-measurement state only with the aid
of the subsequent measurement of an observable $d, a=f(d),$
described by the operator $\widehat{d}$ with nondegenerate
spectrum. We denote by $\{ \phi_{km} \}$ bases in subspaces ${\cal
H}_m =P_m {\cal H}.$ Then by von Neumann
\begin{equation}
\label{GHZ1} \rho^\prime= \sum_m \sum_k  \langle \rho \phi_{km},
\phi_{km} \rangle P_{\phi_{km}}.
\end{equation}
It seems that  experimental investigations to compare  von
Neumann's or L\"uders' laws of transformation of states have never
been performed. It is amazing! It is not so complicated to check
whether after measurement of an observable $a$ (represented by a
self-adjoint operator with degenerate purely discrete spectrum)
the post measurement state is a pure state (L\"uders' postulate)
or some statistical mixture (von Neumann's viewpoint).

\subsection{Conditional probability}

As in the classical Kolmogorov and von Mises probabilistic models,
in  QM Born's postulate about the probabilistic interpretation of quantum states
 should be completed  by a definition
of conditional probability.  We present the contemporary
definition which is conventional in quantum logic   and
quantum information theory\footnote{This definition is based on
L\"uders postulate. Von Neumann's assumption that observable
should have nondegenerate spectrum was totally ignored.} :

\medskip

{\bf Definition.} {\it Let physical observables $a$
and $b$ be represented by self-adjoint operators with purely
discrete (may be degenerate) spectra:
\begin{equation}
\label{PRR1} \widehat{a}=\sum_{m} \alpha_m P_m^a ,\;
\widehat{b}=\sum_{m} \beta_m P_m^b ,
\end{equation}
where $P_m^a$ and $P_m^b$ are projectors on subspaces corresponding
to eigenvales  $\alpha_m$ and $ \beta_m.$

Let $\psi$ be a pure state and let $P_k^a \psi\not=0.$ Then the
probability to get the value $b=\beta_m$ under the condition that
the value $a=\alpha_k$ was observed in the preceding measurement
of the observable $a$ on the state $\psi$ is given by}
\begin{equation}
\label{PRR0} {\bf P}_\psi (b=\beta_m\vert a=\alpha_k) \equiv
\frac{\Vert P_m^b \; P_k^a\psi \Vert^2}{\Vert P_k^a \; \psi
\Vert^2}
\end{equation}

\medskip

Sometimes the symbol ${\bf P}_\psi (P_m^b \vert P_k^a)$ is used.
Set
$$
\psi_k^a = \frac{P_k^a \psi}{\Vert P_k^a\psi \Vert}.
$$
Then
$$
{\bf P}_\psi (b=\beta_m\vert a=\alpha_k)= \Vert P_m^b \psi_k^a
\Vert^2= {\bf P}_{\psi_k^a}(b= \beta_m)
$$
Let $\widehat{a}$ has nondegenerate spectrum. We can write:
$$
{\bf P}_\psi (b=\beta_m\vert a=\alpha_k)= \Vert P_m^b e_k^a \Vert^2
$$
(here $\widehat{a} e_k^a = \alpha_k e_k^a).$ Thus the conditional
probability in this case does not depend on the original state
$\psi.$ We can say that the memory  about the original state was
destroyed.

If  also the operator $\widehat{b}$ has nondegenerate spectrum
then we have: $$ {\bf P}_\psi (b=\beta_m\vert a=\alpha_k)=
\vert\langle e_m^b,e_k^a \rangle \vert^2
$$ and
$$
{\bf P}_\psi (a=\alpha_k\vert b=\beta_m)= \vert\langle e_k^a,e_m^b \rangle \vert^2.
$$
By using  symmetry of the scalar product we obtain:

 {\it Let both operators
$\widehat{a}$ and $\widehat{b}$ have purely discrete nondegenerate
spectra and let $P_k^a \psi\not=0$ and $P_m^b \psi\not=0.$ Then
conditional probability is symmetric and it does not depend on the
original state $\psi:$}
$$
{\bf P}_\psi (b=\beta_m\vert a=\alpha_k)= {\bf P}_\psi
(a=\alpha_k\vert b=\beta_m)= \vert \langle e_m^b,e_k^a \rangle \vert^2.
$$

We now invent the notion of conditional probability for a quantum
statistical state given by a density operator $\rho.$ Let two
observables be represented by operators (\ref{PRR1}). Then the
probability to get the value $b=\beta_m$ under the condition that
the value $a=\alpha_k$ has been observed in the preceding
measurement of the observable $a$ on the state $\rho$ is given by

\begin{equation}
\label{CCR} {\bf P}_\rho(b=\beta_m\vert a=\alpha_k)= \rm{Tr} \;
\rho_k^a \; P_m^b,\; \;\mbox{where}\; \rho_k^a= \frac{P_k^a
\;\rho\; P_k^a}{\rm{Tr}\; \rho \;P_k^a}.
\end{equation}
Here (according to L\"uders) the density operator $\rho_k^a$
describes the quantum state after the result $a=\alpha_k$ was
obtained. We shall also use the notation ${\bf P}_\rho(P_m^b \vert
P_k^a)$ for  ${\bf P}_\rho(b=\beta_m\vert a=\alpha_k).$

We remark that the validity of the conventional (L\"uders' type) definition of conditional probability
in QM can be in principle tested experimentally. Unfortunately, such experimental tests have never been
performed.

\end{document}